\documentclass[pra,twocolumn,floatfix,aps,psfig]{revtex4}
\usepackage{graphicx}
\usepackage{dcolumn}
\usepackage{amsmath}
\newcommand{\ee}{\end{equation}}
\newcommand{\bea}{\begin{eqnarray}}
\newcommand{\eea}{\end{eqnarray}}

\begin{document}

\author{Sadhan K. Adhikari$^{1}$\footnote{adhikari@ift.unesp.br;
URL: www.ift.unesp.br/users/adhikari} 
and Mahir S. Hussein$^{2,3}$
\footnote{Martin Gutzwiller  Fellow, 2007/2008}}
\affiliation{$^1$Instituto de F\'isica Te\'orica, UNESP $-$ S\~ao Paulo 
State University, 01.405-900, SP, Brazil \\ 
$^{2}$Max-Planck-Institut f\"ur Physik komplexer Systeme,  
D-01187 Dresden, Germany \\
$^{3}$Instituto de F\'{i}sica, Universidade de S\~{a}o Paulo,
C.P. 66318, 05315-970 S\~{a}o Paulo, SP, Brazil}
\title{Semi-classical scattering in two 
dimensions}

\begin{abstract} The semi-classical limit of quantum-mechanical
scattering in two dimensions (2D) is developed. 
We derive the Wentzel-Kramers-Brillouin and Eikonal results for 2D 
scattering. 
No backward
or forward glory scattering is present in 2D. Other phenomena, such as
rainbow or orbiting do show up. 
\end{abstract}

\maketitle

\section{Introduction}
\label{I}
The quantum mechanical scattering problem in two space dimensions (2D) 
has
been addressed Refs. \cite{lap, adh,bar}. 
The  inherent circular symmetry in 2D   renders the 
free
partial-wave radial function to be the ordinary Bessel (regular at
the origin) and Neumann (irregular) functions, in 
contradistinction to
the the three-dimensional (3D) case where these functions become the 
spherical counterpart 
(spherical Bessel and Neumann functions). 
2D
scattering has application in surface physics, besides being an
interesting pedagogic topic at the
graduate level.
 
The semi-classical (SC) limit
of the 3D scattering addressed in Refs. \cite{sc,newton}  is an 
important subject  as it shows the connection between quantum 
and
classical mechanics in a particularly transparent form.
It has also important application in
atomic, molecular and nuclear scattering where, at large energies,  one 
usually 
encounters
sums over prohibitively large numbers of partial
waves. Such sums are converted into integrals and 
efficiently evaluated in SC
 scattering. Further, the use of semi-classical ideas is a common 
practice in
the description of wave optics phenomena such as the meteorological
rainbow and glory \cite{LF01}. The concepts employed in the wave optics 
field found a very natural
adaptation in SC 3D  scattering, where one deals with the
optics of matter waves. 

It would be only natural to extend the SC limit
to 2D scattering, which should find  interesting application in surface 
physics  where the results of 2D quantum scattering are often 
applied. Here we develop the SC version of 2D
scattering using the deflection function \cite{newton} approach.
The SC  limit covers a situation where the potential varies very little 
over a distance of the order of de Broglie wave length. The potential 
$V$ need not be weak so long as energy $E\gg |V|$. Hence the domain of 
validity is different from that of quantum Born approximation 
\cite{sakurai}. Under these situations the semiclassical path concept is 
valid and the quantum wave-function is replaced by the semiclassical
  Wentzel-Kramers-Brillouin (WKB) wave function \cite{newton,sakurai}.

Here we
develop the SC version of 2D scattering using the deflection function 
\cite{newton,hus} approach. We present results for 2D scattering in the  
  WKB and Eikonal approximations \cite{sakurai} as well as a
comprehensive discussion of rainbow and glory scattering and orbiting. 
From the asymptotic form of 2D SC wave function we extract the phase
shift and construct the Eikonal result for scattering (cross section)
using this phase shift. The same Eikonal result can also be obtained
in 2D 
from an evaluation of the quantum Born approximation to scattering
amplitude using the SC WKB wave function, as has been done in 3D 
\cite{sakurai}. We leave this as an exercise for the reader,
which can be performed in a straightforward fashion following the 3D
analysis of Ref. \cite{sakurai}. 

In
Sec. \ref{II}, we give a brief account of 2D quantum scattering 
following Ref. \cite{adh}. 
In
Sec. \ref{III} a self-contained discussion of SC 2D 
scattering is presented.  Specifically, we present results for 
WKB wave function and phase shift, Eikonal phase shift and scattering 
amplitude, SC cross section, and rainbow cross section. 
In Sec. \ref{IV}, a summary of our findings is
 given. In Appendix A we present a study of  Rainbow 
scattering cross section at different angles where we also present a 
project for the reader. In Appendix B we have left an exercise.

\section{Quantum Scattering in 2D} \label{II}


The time-independent Schr\"odinger equation for reduced mass $\mu$, 
potential $V({\bf r})$, 
energy $E$, describing scattering is
\begin{equation}\label{1eq}
-\frac{\hbar^2}{2\mu}\nabla^2\psi({\bf{r}}) + V\psi({\bf{r}}) = 
E\psi({\bf{r}}).
\end{equation}
In polar coordinates [${\bf{r}}\equiv (r,\theta)$], Eq. (\ref{1eq})
becomes \cite{lap,adh}
\begin{equation}\label{2eq}
\frac{1}{r}\frac{\partial}{\partial 
r}\left(r\frac{\partial\psi}{\partial 
r}\right)+ \frac{1}{r^2}\frac{\partial^2\psi}{\partial^2\theta} +[k^2 
- U(r)] \psi =0,
\end{equation}
where $k^2 = {2 \mu E}/{\hbar^2}$ and $U(r) = {2 \mu V(r)}/{\hbar^2}$. 
Here $U(r)$ is a central potential (depending only on the magnitude of 
$\bf{r}$).
Equation (2) can be  solved using the separation of variables 
method \cite{lap,adh},
namely, one writes $\psi({\bf{r}})=\sum_m R_m(r)T_m(\theta)$ with  
${d^2T_m(\theta)}/{d\theta^2} = - m^2
T_m(\theta)$, so that $T_{m} (\theta) =
\pi^{\frac{1}{2}} \cos(m\theta)$, thus yielding
\begin{equation}\label{3eq}
\frac{1}{r}\frac{\partial}{\partial r}r\frac{\partial R_{m}(k, r)}
{\partial r} + \left[ k^2 - U(r) - \frac{m^2}{r^2}\right] R_{m}
(k, r)= 0.
\end{equation}
Introducing $\hat R_{m} \equiv  {{R}_{m}} 
{r^{\frac{1}{2}}}$, Eq.
(\ref{3eq})  can be rewritten as
 \begin{eqnarray}\label{4eq}
 \frac{d^2 \hat{R}_{m}(k, r)}{dr^2} +  F(r)\hat{R}_{m}(k, r) 
= 0, \\
F(r)\equiv \left[k^2 - U(r)
- \frac{(m^2 - \frac{1}{4})}{r^2} \right]. \label{4eq1}
\end{eqnarray}
The function $\hat{R}_{m}(t,r)$ should 
be a regular at $r=0$ with  
asymptotic form similar to that of the free
solution (no potential). Setting $U(r)$ equal to zero in Eq. 
(\ref{3eq}), one gets the ordinary Bessel equation, whose solutions are 
the regular
Bessel  [$J_{m}(kr)$] and irregular Neumann [$N_{m}(kr)$] 
functions, 
with asymptotic forms for $kr\gg 1$ \cite{newton}:
\bea\label{5eq} 
J_{m}(kr) \rightarrow \left(\frac{2}{\pi kr}\right)^{\frac{1}{2}} \cos 
\left[ kr 
-\frac{m\pi}{2}-\frac{\pi}{4}\right],\\
\label{6eq}
N_{m}(kr) \rightarrow \left(\frac{2}{\pi kr}\right)^{\frac{1}{2}} \sin 
\left[ kr -\frac{m\pi}{2}-\frac{\pi}{4}\right].
\eea
The asymptotic form of radial  function $\hat R_{m}(kr)$, being
a linear combination of $J_{m}(kr)$ and $N_{m}(kr)$, is taken as 
\cite{adh} 
\begin{equation}\label{7eq}
\hat{R}_{m}(k, r) \rightarrow A_{m} (k)^{-\frac{1}{2}} \cos \left[ 
kr -\frac{m\pi}{2}-\frac{\pi}{4} + \delta_m(k)\right],
\end{equation}
where 
$\delta_m(k)$ is  the scattering phase shift and the  
 constant $A_{m}$ is given by
$A_{m} = 2 i^{m} \epsilon_m (2 \pi)^{-\frac{1}{2}} e^{i \delta_m(k)},
$
with $\epsilon_m=2$ for $m\ne 0$ and $\epsilon_0=1$. 

The scattering amplitude is written in terms of 
$\delta_m(k)$ as  \cite{adh}
\begin{equation}\label{9eq}
f(k,\theta)=\sqrt{\frac{2}{\pi}}\sum_{m =
0}^{\infty} \cos (m \theta)\epsilon_me^ {i\delta_m(k)}\sin \delta_m(k).
\end{equation}
The differential scattering cross section  is 
$d\sigma_{2D}/d\theta
=|f(k,\theta)|^2/k$ and the total cross section is 
$\sigma_{2D}(k^2)
=\int_0^{2\pi}
(d\sigma_{2D}/d\theta)
d\theta = 4 
\sum_0^{\infty}\epsilon_m\sin^2 \delta_m(k)$. 
In 2D the scattering amplitude has the dimension of L$^{1/2}$
(L in 3D) 
and 
cross section has the dimension of L (L$^2$ in 3D).

\section{Semi-classical  scattering in 2D}
\label{III}
When the local de Broglie  wave length is short compared to the
distance over which the potential changes appreciably, concepts of 
classical scattering can be used to describe quantum scattering. 
The problem is formulated in terms of classical 
quantities, such
as the impact parameter, $b$, and the deflection function, $\Theta
(k^2,b)$ \cite{hus,newton}: the validity 
criterion 
is 
${\delta\theta_{\mathrm{opt}}}/{\Theta(k^2,b)} \ll 1 $
for all $b$
and $E$. Here $\delta\theta_{\mathrm{opt}}$ is the optimal angular
dispersion of the scattering particle around a classical path. 
This corresponds to the high-energy region where many 
angular momenta contribute and angular momentum is often treated as a 
continuous variable and the quantum-mechanical partial-wave sum is 
replaced by an integral.  

The useful functional relation of the deflection function 
$\Theta (k^2,b)$, obtained from the Hamilton-Jacobi equation of motion, 
 in both 2D and 3D  is given by \cite{newton},
\begin{equation}\label{32eq}
\Theta (k^2, b) = \pi - 2bk \int_{r_{0}}^{\infty} 
{dr}
{r^{-2} F^{-\frac{1}{2}}}(r)
\end{equation}
where    $r_{0}$ is the classical 
turning point:  $F(r_0)=0.$

At this stage we recall that Eqs. (\ref{4eq}) and (\ref{4eq1}) can also 
be solved by the SC WKB
approximation \cite{newton} with the ansatz
$\hat R = (-F)^{-1/4}\exp[\int_{r_0}^r dr'(-F)^{1/2}$ 
\cite{newton} near $r=0$. As we cross the turning point and reach 
the asymptotic region we take $(-F)^{1/2}\to iF^{1/2}$, $(-F)^{-1/4}\to 
F^{-1/4}e^{-i\pi/4}$. In this fashion the wave-function is real in both 
regions. Then for $r>r_0$ \cite{newton}
\bea\label{51eq}
&&\hat R = F^{-1/4} \cos\left( \int_{r_0}^rdr'F^{1/2} -\frac{\pi}{4}        
\right),\\
&&\approx_{{r\to\infty}} \cos\left[ 
kr+ \int_{r_0}^\infty dr(F^{1/2}-k)-kr_0 -\frac{\pi}{4}
\right]. \label{52eq}
\eea 
From Eqs. (\ref{7eq}) and (\ref{52eq}), the WKB phase shift becomes
\begin{eqnarray}\label{39eq}
\delta^{\mathrm{WKB}}_{2D}(k,m) &=& m\pi/2+ \int_{r_0}^\infty
dr  (F^{1/2}-k)-kr_0.
\end{eqnarray}
Differentiating Eq. (\ref{39eq}) with respect to $m\equiv kb$  we get 
\bea
\frac{d\delta_{2D}^{\mathrm{WKB}}(k,m)}{dm}= 
\frac{\pi}{2}-m\int_{r_0}^\infty
dr r^{-2}F^{-1/2}. 
\eea
Comparing with Eq. (\ref{32eq}) we establish 
\bea\label{53eq}
\Theta=2 \frac{d 
\delta_{2D}^{\mathrm{WKB}}(k,m)}{dm}.
\eea
At very high energies many angular momentum contribute and the so-called 
Eikonal approximation can be derived by replacing the $m$-sum in Eq. 
(\ref{9eq}) by an integration with the identification $m=kb$ 
\cite{newton}: 
\begin{equation}\label{41eq}
f_{2D}^{\mathrm{Eik}}(k,b)= \frac{-ik}{(2\pi)^{1/2}}\int_0^\infty db 
\cos(kb\theta) \biggr[ e^{2i \delta_{2D}^{\mathrm{Eik}}}-1   
\biggr].
\end{equation}
The Eikonal phase shift $\delta_{2D}^{\mathrm{Eik}}$ is now obtained 
from 
Eq. (\ref{39eq}) by expanding $F$ in a power series in $U(r)/k^2$,  
setting $r_0=b$, $m=kb$ and $m^2-1/4 \approx k^2b^2$, 
and keeping the lowest order term:
\cite{newton}
\bea  \label{42eq}
\delta_{2D}^{\mathrm{Eik}}= \frac{-1}{2k}\int_b^\infty \frac{r dr 
U(r)}{(r^2-b^2)^{1/2}}.
\eea
Incidentally, in 3D the Eikonal approximation  is similar to 
Eqs. (\ref{41eq}) and (\ref{42eq}) except for the replacement 
$db\to (2\pi)^{1/2} bdb$, $\cos(kb\theta)\to J_0(kb\theta)$ 
\cite{newton}. 

We now recast the
scattering amplitude into a sum of integrals \cite{newton} using  the 
Poisson sum formula \cite{MF}:
\begin{equation}\label{17eq}
\sum_{\nu = -\infty}^{\infty} F_{\nu} = \sum_{\kappa = 
-\infty}^{\infty}e^{ -i\pi  \kappa}\int_{-\infty}^{\infty} dy F(y) 
e^{2i\pi \kappa y}. 
\end{equation}
Formula (\ref{17eq})
can be applied to  scattering amplitude (\ref{9eq}) once
$\cos(m\theta)$ is expressed in terms of exponentials:
\begin{equation}\label{18eq}
f(k, \theta) = f^{(+)}(k, \theta) + f^{(-)}(k, \theta)
\end{equation}
where the amplitudes $f^{(+)}(k, \theta)$ and $f^{(-)}(k, \theta)$ 
correspond to the $\exp(im\theta)$ and $\exp(-im\theta)$ branches of the 
cosine function, respectively. Then using Eq. (\ref{17eq})
\begin{eqnarray}\label{19eq}
&&f^{(\pm)}(k, \theta) = \frac{1}{\sqrt{ 2 \pi}} \sum_{\kappa = 
-\infty}^{\infty}e^{- i\pi \kappa }  \nonumber \\ 
&\times& \int_{0}^{\infty} \exp \big[ i \{ 2 
\delta(k,m)+  2 \pi  \kappa m  \pm  m \theta \} 
\big] dm,
\end{eqnarray}
where now the discrete sum over angular momentum is replaced
by an integration and the SC phase-shift $\delta(k,m)$ is essentially 
the WKB phase shift. (We drop the label WKB in the following.)   

The integrals in Eq.  (\ref{19eq}) can be evaluated in the
SC regime of stationary phase approximation (SPA) when the
phase shift, being an action divided over $\hbar$, is very large. The
formula of the SPA is based on evaluating integrals of the type $ I =
\int dm \exp[i \zeta(m)]$, where the phase 
$\zeta(m)\equiv [2\delta(k,m)+2\pi\kappa m\pm m\theta] $ 
is real on the 
line of
integration and analytic in some region surrounding it. If $\zeta(m)$
varies rapidly with $m$, 
the integral will be vanishingly small, except
in cases where $\zeta$ has extrema on the line of integration. 
One thus
expands $\zeta(m)$ around the extremum and keeps terms  up to second 
order 
in $m$. For  one extremum
point, $m_s$ 
\begin{equation}\label{20eq}
\zeta (m) = \zeta (m_s) + \zeta''(m_s) ( m - m_s)^2/2,
\end{equation}
where prime denotes derivative. 
The integral  can now be  performed to yield 
\begin{eqnarray}\label{21eq}
I &\equiv& e^{i\zeta(m_s)} \int _{-\infty}^{\infty} \exp\left[ i 
\zeta''(m_s) ( m - m_s)^2 /2\right] dm \nonumber \\ &=& 
\left[{2\pi 
i}/{\zeta''(m_s)}\right]^{\frac{1}{2}}  e ^{i\zeta(m_s)}, 
\end{eqnarray}
where the limits of integration have been extended to 
$\pm \infty$ since
the contribution of the far away points are strongly suppressed by
oscillation.
The above result can be generalized for several extrema, $m_s^{(j)}$:
\begin{equation}\label{22eq}
I = \sum _{j} \left[{2\pi i}/{\zeta''(m_s^ {(j)})}\right]^
{\frac{1}{2}}  
e^{i\zeta(m_s^{(j)})}
\end{equation}
The power of the SPA is that it allows replacing an infinite sum, such
as the $m$-sum in $f$ by a sum of only  few contributions arising 
from
the stationary points. The Poisson sum can now be performed, by seeking
the stationary points of the total phase $\zeta(m) $ via
$d\zeta(m)/dm=0$:, viz, 
\begin{equation}\label{24eq}
\pm \theta = -2 \frac{d \delta (k,m)}{dm} - 2 \kappa \pi. 
\end{equation}
In Eq. (\ref{24eq}) 
$\kappa$  
represents the 
number of 
times the particle 
circles around the scatterer for an attractive interaction. This 
phenomenon is known as orbiting.   Taking 
$\kappa = 0$ in Eq. (\ref{24eq}) and comparing with Eq. (\ref{53eq})
we find that  the scattering angle $\theta$ is related to the  
deflection function $\Theta: \Theta=\mp \theta$ \cite{newton}, where 
the minus sign refer to $f^{(+)}$ and the plus sign to $f^{(-)}$. If 
$\Theta $ is positive $f^{(-)}$ dominates and {\it 
vice versa.} The case 
$\Theta=0$ corresponds to a situation where $f^{(+)}$ and 
$f^{(-)}$ are 
comparable in size and the cross section should show a 
Fraunhofer-type 
interference.
From Eqs. (\ref{19eq}), (\ref{21eq}), and (\ref{24eq}) we get 
\begin{equation}\label{25eq}
f^{(\pm)}(k, \theta) =  \sum_{m_{s}^{(\pm)}}
e^{
i ( \pm m_{s}^ {(\pm)}\theta +2\delta + {\pi}/{4})} 
 {\left[ 
\frac {d 
\Theta}{d m}\right]}  ^{-\frac{1}{2}},
\end{equation}
where the $m$-derivative is evaluated at the stationary points 
$m_s^{(\pm)}$.
The differential cross section  now 
is
\begin{eqnarray}
\frac{d\sigma_{2D}}{d \theta}
& =& \frac{1}{k} [ |f^{(+)} + f^{(-)}|^2] \\ &=& 
\frac{1}{k} 
\sum_{m_s^{(\pm)}} 
\biggr[
\frac{d \sigma_{2D}^{\mathrm{Classical}}}{d 
\theta}\biggr]_{m_{s}^{(\pm)}}
+ I^{(\pm)} ( k^2, \theta) ,
\label{26eq}
\end{eqnarray}
where  
$(d \sigma_{2D}^{\mathrm{Classical}}/d \theta)_{m_{s}^{(\pm)}}$
is the 
classical differential cross section arising from 
stationary point $m_{s^{(\pm)}}$
\begin{equation}\label{27eq}
\biggr[\frac{d \sigma_{2D}^{\mathrm{Classical}}}{d
\theta}\biggr]_{m_{s}^{(\pm)}}
= \frac{1}{k} 
\left| 
\frac{d \Theta}{d m}\right|^{-1}
\end{equation}
and the interference terms $I^{(\pm)} ( k^2, \theta)$ are associated 
with 
the two branches of the amplitudes.

If only one stationary point contributes to, say,   $f^{(-)}$, 
then
the SC  differential scattering cross section is just the
classical one (no interference term). This is similar to the 3D Coulomb
scattering \cite{newton} where the exact quantum mechanical cross 
section is identical
to the classical one, as the deflection function $\Theta(m_{s})$ is a
monotonic function of the classical impact parameter, identified as
$b=km_{s}$. In the 2D case, this  identification is just the same and 
 one finally obtains for the
2D classical differential cross section
\begin{equation}\label{28eq}
\frac{d \sigma_{2D}^{\mathrm{Classical}}}{d
\theta}
= \left|\frac {d 
\Theta}{d 
b}\right|^{-1}
\end{equation}
to be compared to the classical 3D differential cross section 
\cite{newton}
\begin{equation}\label{29eq}
\frac{d\sigma_{3D}^{\mathrm{Classical}}}{d\Omega} = \frac{b}{\sin\theta} 
\left|\frac{d 
\Theta}{db}\right|^{-1} 
\end{equation}
If the differential cross sections
are written as differentials with respect to $\theta$,  ($d\Omega = 
2\pi \sin \theta d\theta$,) we have from Eq. (\ref{29eq}) 
\cite{newton}
\begin{equation}\label{30eq}
\frac{d\sigma_{3D}^{\mathrm{Classical}}}{d\theta} = 2 \pi b 
\left|\frac{d b}{d\theta}\right| 
= \frac{d (\pi b^2)}{d\theta}
\end{equation}
to be compared to the 2D cross section (\ref{28eq}), namely 
\begin{equation}\label{31eq}
\frac{d\sigma_{2D}^{\mathrm{Classical}}}{d\theta} = \left|\frac {d b}{d 
\theta}\right|
\end{equation}
where in Eqs. (\ref{30eq}) and (\ref{31eq}) we have 
set $|\Theta| =|\theta|$ 
as the 
stationary point condition 
requires. Correspondingly, the total cross sections integrated over all 
angles are $\sigma_{3D}^{\mathrm{Classical}}=\pi b^2_{\mathrm{max}}$ and 
$\sigma_{2D}^{\mathrm{Classical}}=2 b_{\mathrm{max}}$, where the 
integration is limited to a maximum $|b|=b_{\mathrm{max}}. $

From Eq. (\ref{28eq}) we see that when the deflection function is 
stationary with respect to variations of impact parameter $b$, e.g. 
$d\Theta/db=0$, the differential cross section is infinite. Moreover, 
Eq. (\ref{32eq}) shows that  $\Theta(b)$ will not be a monotonic 
function of $b$ and 
the differential  cross section may exhibit interference like in a 
rainbow \cite{newton}.    
Here angular deflection, as measured by the deflection 
function, can be positive or negative. The experimental scattering angle 
$\theta$ is limited by $0<\theta<\pi$. Hence the values of impact 
parameter $b$, for which  $\Theta>0$,  contributes more to $f^{(-)}$ and 
negligibly 
to $f^{(+)}$ and {\it vice versa.} If $d\Theta/db=0$ for 
positive 
$\Theta,$ 
$f^{(-)}$ will show rainbow characteristics and {\it vice 
versa.} 
If deflection function $\Theta(m)$, with 
$m=bk$,  is expanded 
about the 
rainbow value $m_r$ of angular momentum $m_r \equiv kb_r$ at which 
$\Theta'(m_r)=0$, 
\begin{equation}\label{app}
\Theta(m)=\Theta(m_r)+\Theta''(m_r)(m-m_r)^2/2+..., 
\end{equation}
then 
the phase $\zeta $ of Eq. (\ref{19eq}) becomes 
$\zeta_\pm=2\delta(m_r)+\Theta(m_r)(m-m_r)+\Theta''(m_r)(m-m_r)^3/6+2\pi\kappa 
m\pm m\theta,$ [compare with Eq. (\ref{20eq})]. (For repulsive 
potential there is no orbiting and $\kappa$ should be set equal to 
zero. However, the results for amplitude and cross section below do not 
depend on $\kappa$.) 
Then the resulting integral in Eq. (\ref{19eq}) has the same form as in 
Eq. (\ref{21eq}) and can be expressed in terms of the Airy function  
\cite{newton,MF}  
$\mbox{Ai}(x)=(1/2\pi)\int_{-\infty}^{\infty}\exp[i(tx+t^3/3)]dt$, 
leading to 
\bea\label{55eq}
|f(k,\theta)|= 
\frac{\sqrt{2\pi}}{[\Theta''(m_r)/2]^{1/3}}\mbox{Ai}
\left[\frac{\theta-\theta_r}{[\Theta''(m_r)/2]^{1/3}}  
\right]
\eea
 and the cross 
section is 
\begin{equation}\label{37eq}
\frac{d\sigma_{2D}}{d\theta} = \frac{{2^{5/3}k^{1/3}\pi}} 
{({|d^2\Theta/db^2|_{b_r}})^{\frac{2}{3}}} \mbox{Ai}^{2} 
\left[ 
\frac{2^{1/3}k^{2/3}(\theta 
-\theta_{r})}{({|d^2\Theta/db^2|_{b_r}})^{\frac{1}{3}}} \right]
\end{equation}
where $\theta_{r}\equiv \Theta(m_r)$ is the rainbow angle. The Airy 
function being an oscillating function the rainbow-like interference 
pattern in the differential cross section (\ref{37eq}) is confirmed.
For $\theta < \theta_{r}$, the expression (\ref{37eq}) 
oscillates and 
the particle is scattered 
into the
illuminated region. For $\theta >  \theta_{r}$, the Airy function  dies 
out rapidly, as the particle scatters into the dark side of the rainbow.
 This feature is of a purely
 quantum nature related to the phenomenon of tunneling; classically,
 there is no scattering for $\theta > \theta_r$, and thus the name
 "dark". Quantum mechanics supplies some "illumination" in this
 classically forbidden region.



Besides the
rainbow, one has the effect of glory scattering in 3D, both near $\theta 
= 0$
(forward glory) and near $\theta = \pi$ (backward glory). For example, 
when back scattering is possible for impact parameter other than zero 
($b\ne 0$), 
one has $\theta = \pi$ and cross section (\ref{29eq}) can be infinite 
due to the vanishing of $\sin \theta$. It occurs whenever $\Theta$ 
of Eq. (\ref{32eq}) goes smoothly through 0 or through an integral 
multiple 
of $\pi$ \cite{newton}. In 2D cross section (\ref{28eq}) the $\sin 
\theta$ term is absent and one cannot have glory scattering.
However, one can obtain oscillating cross section near $\theta =0$ or 
$\pi$ owing to near equality of $f^{(+)}$ and $f^{(-)}$ [described by 
Bessel function(s)].

The
phenomenon of orbiting  
will take place in both 2D and 3D when the effective potential $U(r)
+ {b^2k^2}/{r^2}$ has a maximum for a certain $b_{0}$ and the energy 
of
the particle coincides with this maximum \cite{newton}. 
One has to verify whether 
${d ( U + {b^2k^2}/{r^2})}/{dr} = 0$ for a certain  $b_{0}$ 
and $r_{0}$ and check if $k^2 = U(r_{0}) + k^2{b_{0}^2}/{r_{0}^2
}$. 
From  
expression (\ref{32eq}) for
$\Theta$ , one can show that in the case of orbiting \cite{newton}
\begin{eqnarray}\label{38eq}
\Theta (k^2, b) &=& \mbox{const} + c \ln [ ({b - b_{0}})/{b_{0}}], \quad  
b 
> 
b_{0} \\
\label{38eqa}
\Theta (k^2, b) &=& \mbox{const} + 2 c \ln [ ({b_{0} - 
b})/{b_{0}}],\quad  
b 
< 
b_{0}.
\end{eqnarray}
As $b\to b_0,$ $\Theta (k^2, b)$ can be infinitely large and for 
attractive 
interaction the particle can go around the center many times. Then the 
differential cross section made from  many components $-$ one 
from each orbit $-$ could be greatly different   from the classical 
results (\ref{28eq}) and (\ref{29eq}) and will show interference 
effect \cite{newton}.

The interference terms of
the SC  2D cross 
section (\ref{26eq})
is absent in cross section (\ref{28eq}) for a single stationary point 
$n_s$. 
The case of two stationary points in Eq. (\ref{26eq}) leads to 
an interference term and 
deserves
 special attention. We assume that  
deflection function $\Theta(k^2, b)$ is a double-valued function. For a 
given $\theta$, there are two impact parameters, $b_{1}$, 
$b_{2}$ ($n_1=kb_1$ and $n_2=kb_2$), that contribute to Eq. 
(\ref{25eq}). If there are two contributions to , say, $f^{(-)}$ in Eq. 
(\ref{25eq}),
then $ d\Theta/dm$ will be positive at one of these points, and negative
at the other. This results in an extra phase of $\pi/2$ in the
contribution to the amplitude in Eq. (\ref{25eq})
coming from the latter stationary point.
Then from Eq. (\ref{25eq}) the scattering 
cross 
section  can be written 
as
\begin{eqnarray}\label{33eq}
&& \frac{d\sigma_{2D}^{\mathrm{SC}}}{d\theta} =  \left|\frac{d 
b_{1}}{d\theta}\right|+ \left|\frac{d 
b_{2}}{d\theta}\right| + 
2\left|\frac{d b_{1}}{d\theta} \frac{d 
b_{2}}{d\theta}\right|^{\frac{1}{2}}\nonumber \\ &\times& \sin [k 
(b_{1} -b_{2})\theta + 2 \{
\delta(b_{1}) - \delta(b_{2})\} ].
\end{eqnarray}
Equation (\ref{33eq}) clearly shows the wave nature of the scattering
process. The interference term involving the sine term   is  purely 
quantum 
mechanical
would be absent in classical scattering.
The classical scattering cross section 
is an incoherent sum of contributions. The
interference pattern has an angular period, $P_{\theta}$ (angular
distance between two adjacent maxima in the scattering cross section), 
which can
be read off from the above, as
$P_{\theta} = {2\pi}/{|k(b_{1} - b_{2})|}.$
When $b_{1}$ approaches $b_{2}$,  the analysis leading to Eq. 
(\ref{33eq})  
breaks 
down and one has to resort to a more elaborate treatment, called the 
Uniform Approximation \cite{CFU}, which provides a generalization of 
Eqs. (\ref{55eq}) and (\ref{37eq}) involving the Airy function and its 
derivative.




\section{summary}
\label{IV}
We have presented a comprehensive discussion of semiclassical (SC)
scattering theory in two dimensions.  We have presented  and discussed 
classical and SC cross section in 2D,
WKB and Eikonal approximations. Although, rainbow scattering and 
orbiting take place in 2D, there is no forward and backward glory 
scattering.

\acknowledgments

The work was partially supported by CNPq and FAPESP of Brasil.

\section*{APPENDIX}

\subsection{An Application of Rainbow Scattering}

The semiclassical cross section (\ref{37eq}) exhibits the essential 
quantum
effects of tunneling and interference. To see this more clearly
 we use the asymptotic form of  the Airy function Ai$(x)$  
as $x\to \pm \infty$ \cite{Abram}
\begin{eqnarray}\label{38}
{\mbox{Ai}}(x) &=& \frac{ \exp(
 - 2 x^{3/2}/3)  }{2\sqrt \pi x^{1/4}},\quad x>0, \\
{\mbox{Ai}}(x) &=& \frac{\sin [
2(-x)^{3/2}/3 + \pi/4]  }{\sqrt\pi (-x)^{1/4}} 
,\quad x<0. \label{39}
\end{eqnarray}
Using the asymptotic form (\ref{38}), we find that  the cross section 
(\ref{37eq}) exhibits a decay in angle that goes like 
\begin{equation}
\frac{d\sigma_{2D}}{d\theta}= \frac{ \exp\left[-\frac{4k\sqrt 2 }{3}
\frac{(\theta - \theta_r)^{3/2}}{(|\frac{d^2 \Theta}{d
b^2}|_{b_{r}})^{1/2}}\right]
}
{\sqrt 2(|\frac{d^2\Theta}{db^2}|_{b_r})^{1/2}\sqrt{
\theta - \theta_r}
} ,
\end{equation}
in the classically forbidden angle region, $\theta > \theta_r$, 
which is reminiscent of tunneling effect [classically, there is zero 
scattering in this dark side of the rainbow, see Eq. (\ref{28eq})]. 


In the classically allowed, illuminated, region, $\theta < \theta_r$,
the cross section exhibits interference arising from the contributions
of the two stationary phase points. Using the asymptotic form (\ref{39})
in Eq. (\ref{37eq}), the oscillation in cross section is given by 
 \begin{equation} \label{41} 
\frac{d\sigma_{2D}}{d\theta}= \frac{2\sqrt 2 
\sin^{2}\left[
\frac{2k\sqrt 2}{3} \frac{(\theta_r - \theta)^{3/2}}{|\frac{d^2
\Theta}{d
b^2}|_{b_{r}}^{1/2}}+ \frac{\pi}{4} \right]
}{(|\frac{d^2\Theta}{db^2}|_{b_r})^{1/2}\sqrt{\theta_r-\theta}}.  \end{equation}
One then finds the period of these oscillations as follows. Call the
argument of the $\sin^2$ function above, $g(\theta) = \frac{2k\sqrt
2}{3} \frac{(\theta_r - \theta)^{3/2}}{(|\frac{d^2 \Theta}{d
b^2}|_{b_r})^{1/2}}+ \frac{\pi}{4} $. Suppose at $ \theta = \theta_1 <
\theta_r $, we have a maximum in cross section implying
$\sin^2[g(\theta_1)]=1$, and $g(\theta_1) = \pi/2$. At a slightly 
smaller
angle, $\theta=\theta_2 < \theta_1$, one has another maximum implying
$\sin^2 [g(\theta_2)] = 1$ corresponding to $g(\theta_2) = \pi/2 + \pi$. 
To obtain an estimate of the local period of oscillation $P_\theta$ of
cross section (\ref{41}), defined by $P_\theta\equiv \theta_1-\theta_2$,
we expand $g(\theta)$ near $\theta=\theta_1$ as
$g(\theta)=g(\theta_1)+(\theta-\theta_1)g'(\theta_1)$. Consequently, the
period $P_\theta$ at $\theta=\theta_1$ is $P_{\theta_1}=
-\pi/g'(\theta_1)$. Hence the generic local and angle-dependent period
for any $\theta< \theta_r$ is \begin{equation}\label{42} P_{\theta}
\equiv -\frac{\pi}{\frac{dg(\theta)}{d\theta}} = \frac{\pi}{{k\sqrt 2}
\frac{(\theta_r - \theta)^{1/2}}{(|\frac{d^2 \Theta}{d
b^2}|_{b_r})^{1/2}}}.  \end{equation} 
Thus the smaller the angle
$\theta$ compared to the rainbow angle $\theta_r$, the smaller will be
the period. 

The angle $\theta=\theta_1$ is called an Airy maximum;  in the 
meteorological 
rainbow, it is 
the last maximum in the Airy function, just below $\theta =\theta_r$, 
and 
corresponds to the primary bow, whereas the other maxima are the
supernumeraries. The second maximum at an angle slightly smaller than
that of the primary bow is called the secondary bow. One usually sees
both of these bows in the clear sky after rain, the secondary bow being
quite dim with the order of colors inverted compared to the more
familiar primary bow. 

The period formula (\ref{42}) is an approximate one valid
under condition (\ref{app})
 used in deriving the rainbow
scattering (\ref{37eq}). Keeping only two terms in Eq. (\ref{app}), one 
obtains the solution for two turning points $m_1$ and $m_2$ as 
$m_{1(2)}=m_r\pm\sqrt 
2{\sqrt{\theta_r-\theta}}/{|\Theta''(m_r)|^{1/2}}$ (recall, 
$\Theta(m_r)=\theta_r$, $\Theta(m)=\theta$,
 and $\theta_r>\theta$). Thus the period of the 
Airy$^2$ function in Eq. (\ref{37eq}) is $P_\theta=2\pi/|m_1-m_2|,$ 
which is 
the same as that derived in  Eq. (\ref{42}) (note that $m=kb$).

The correct period should be obtained by
staying at angles much smaller than the rainbow angle [maximum of
$\Theta(m)$] and using  the semiclassical approximation employed in 
obtaining 
Eq. (\ref{25eq})
and the 
quadratic approximation of the phase $\zeta$, given by Eq. 
(\ref{20eq}). [In the semiclassical approximation leading to
Eq. (\ref{25eq}), Eqs. (\ref{21eq}) and (\ref{22eq}) diverge at the 
rainbow caustic. The amplitude  (\ref{55eq}) is free from that 
divergence but the period of oscillation (\ref{42}) obtained from this 
amplitude is not quite right.]
The uniform approximation of Ref. \cite{CFU} combines the good 
features of both approaches and produces the correct period of 
oscillation. However, the algebra is quite involved and without proof we 
quote the result for the angle-independent period so obtained: 
$P_{\theta} = {2\pi}/{|m_1 - m_2|}$,
where $m_1$ and $m_2$ are the $m$ values at two stationary points. 
The period (\ref{42}) can be shown to be  equivalent to  
the result so obtained 
after using the 
relations $m_1 = kb_1$ and $m_2 = kb_2$. (We leave this as a project 
for an advanced reader.)


\subsection{An Exercise for Reader}


We consider an interaction of the form
\begin{eqnarray}
V(r) &=& \frac{A}{r}, \quad r > R_{c},\\ 
V(r) &=& \frac{A}{2R_{c}}\biggr[ 3 - 
\biggr(\frac{r}{R_{c}}\biggr)^{2}\biggr], \quad r < R_{c},
\end{eqnarray}
where $A$ and $R_c$ are constants.
The above potential should not be confused with the Coulomb interaction
between charged particles, which in 2D behaves as ln(r). The deflection
function for the above interaction shows a rainbow maximum and exhibits
the phenomenon of orbiting. The condition for rainbow is that the energy
$E > {3A}/{(2R_{c})}$. The calculation of the deflection function 
is
presented in \cite{hus}.  We leave the  calculation of 
2D SC scattering  with this potential as an exercise.


\begin{thebibliography}{101}

\bibitem{lap} I. R. Lapidus,  ``Quantum-mechanical scattering in 2 
dimensions",
Am. 
J. Phys. {\bf 50}, 45-47 (1982); ``Scattering by 
two-dimensional circular barrier, hard circle, and delta-function ring 
potentials", Am.
J. Phys. 
{\bf 54}, 459-461 (1986); M. J. Moritz and H. Friedrich, 
 ``Scattering by a Coulomb field in two dimensions", Am. J. Phys. {\bf 
66}, 274-274 (1998).
\bibitem{adh} S. K. Adhikari, ``Quantum scattering in 2
dimensions", Am. J. Phys. {\bf 54}, 362-367 (1986).
\bibitem{bar}G. Barton,  ``Rutherford scattering in 2 dimensions",
Am. J. Phys. {\bf 51}, 420-422 (1983);
P. A. Maurone and T. K. Lim, ``More on two-dimensional scattering", 
Am. J. Phys. {\bf 51}, 856-857 (1983).

\bibitem{sc} M. J. Moritz, ``Semiclassical methods in the long wave 
limit", Am. J. Phys. {\bf 70}, 663-663 (2002); 
D. Singleton,  ``Electromagnetic angular momentum and quantum 
mechanics",  Am. J. Phys. {\bf 66}, 697-701 (1998); B. R. 
Holstein,  Semiclassical treatment of above-barrier scattering", 
Am. J. Phys. {\bf 52}, 321-325 (1984).
\bibitem{newton} R. G. Newton, 
\emph{Scattering Theory of Waves and Particles, 2nd Edition,} 
(Springer-Verlag, New York, 1982).
\bibitem{LF01} R. L. Lee and A. B. Fraser,
{\it The Rainbow Bridge: Rainbows in Art, Myth and Science},
(The Pensylvania State University Press, University Park, 2001).
\bibitem{sakurai}J. J. Sakurai, {\it Modern Quantum Mechanics}, 
(Addison-Wesley, Reading, 1994),  pp 392-394.
\bibitem{hus}  M. S. Hussein, Y. T. Chen and F. I. A. Almeida, 
``The validity of the classical description of nuclear scattering", 
Am. 
J. Phys. {\bf 52}, 650-653 (1984).
\bibitem{MF} P. M. Morse and H. 
Feshbach, \emph{Methods of Theoretical Physics}, (McGraw-Hill, New York, 
1953).
\bibitem{CFU} M. V. Berry, ``Uniform approximation for potential 
scattering involving a rainbow", Proc. Phys. Soc. London, {\bf 89}, 
479-490 
(1966).

\bibitem{Abram} M. Abromowitz and I. Stegun, {\it Handbook of 
Mathematical Functions}, (Dover, New York, 1970).

\end{thebibliography}
\end{document}